\documentclass[a4paper]{article}
\usepackage{Odyssey2022}
\usepackage{epsfig,amssymb,amsmath}
\usepackage{multirow}
\usepackage{float}
\usepackage{hyperref}
\hypersetup{colorlinks=true}
\ninept

\newcommand\mat[1]{\mathbf{#1}}
\newcommand\textgreen[1]{\textcolor[rgb]{0,0.5,0}{\textbf{#1}}}

\newcommand\Ian[1]{\textcolor[rgb]{.2,0.2,.7}{[\textbf{Ian}]}}

\title{Closing the Gap between Single-User and Multi-User VoiceFilter-Lite}

\name{Rajeev Rikhye, Quan Wang, Qiao Liang, Yanzhang He, Ian McGraw}

\address{Google LLC, USA \\
{\small \texttt{
\{\href{mailto:rvrikhye@google.com}{rvrikhye},
\href{mailto:quanw@google.com}{quanw},
\href{mailto:wildstone@google.com}{wildstone},
\href{mailto:yanzhanghe@google.com}{yanzhanghe},
\href{mailto:imcgraw@google.com}{imcgraw}\}@google.com}
} }
\begin{document}
\maketitle

\begin{abstract}
VoiceFilter-Lite is a speaker-conditioned voice separation model that plays a crucial role in improving speech recognition and speaker verification by suppressing overlapping speech from non-target speakers. However, one limitation of VoiceFilter-Lite, and other speaker-conditioned speech models in general, is that these models are usually limited to a single target speaker. This is undesirable as most smart home devices now support multiple enrolled users. In order to extend the benefits of personalization to multiple users, we previously developed an attention-based speaker selection mechanism and applied it to VoiceFilter-Lite. However, the original multi-user VoiceFilter-Lite model suffers from significant performance degradation compared with single-user models. In this paper, we devised a series of experiments to improve the multi-user VoiceFilter-Lite model. By incorporating a dual learning rate schedule and by using feature-wise linear modulation (FiLM) to condition the model with the attended speaker embedding, we successfully closed the performance gap between multi-user and single-user VoiceFilter-Lite models on single-speaker evaluations. At the same time, the new model can also be easily extended to support any number of users, and significantly outperforms our previously published model on multi-speaker evaluations.
\end{abstract}

\section{Introduction}
\label{sec:intro}
Speaker-conditioned speech models are a class of speech models that are conditioned on a target speaker embedding, allowing the model to produce personalized outputs. For example, in personalized speaker separation, prior knowledge of a target speaker’s voice profile is used to suppress overlapping speech from non-target speakers ~\cite{Wang2019, Wang2020,wang2018deep,zmolikova2017speaker,vzmolikova2017learning,delcroix2018single,xu2020spex}. In personalized Automatic Speech Recognition (ASR), a speaker’s voice profile is used to improve the overall recognition accuracy ~\cite{he2018streaming, bellpasr,denisov2019end,shi2021improving}. Additionally, in personalized Voice Activity Detection (VAD), the target speaker profile is used to determine when the target speaker begins or stops talking, which in turn improves the accuracy of downstream  components such as ASR~\cite{ding2019personal}.

While beneficial, speaker-conditioned speech models are often only limited to a single enrolled user. This makes them incompatible with many devices, such as smart displays and smart speakers, which currently support multiple users~\cite{multiuser}. One naive approach to mitigate this would be to have multiple passes of the same model --- one pass for each enrolled user. This approach is, however, computationally expensive and unacceptable for on-device applications. As such, extending speaker-conditioned models to support multiple users remains an open and relevant problem \cite{kanda2020joint, han2020continuous}

To overcome this limitation, we previously described an attention-based~\cite{vaswani2017attention} speaker selection mechanism, and extended VoiceFilter-Lite to support an arbitrary number of enrolled users ~\cite{rikhye2021multiuser} (see Fig.~\ref{fig:model-arch}). This multi-user VoiceFilter-Lite model significantly reduces speech recognition Word Error Rate (WER) and speaker verification Equal Error Rate (EER) when the input audio contains overlapping speech. We also demonstrated how this multi-user VoiceFilter-Lite model is critical to a personalized keyphrase detection system on shared devices ~\cite{rikhye2021personalized, rikhye2021multiuser} by reducing the false rejection rate caused by speaker mis-identification. Although the performance is promising, this original multi-user model suffered from two issues. First, on single-user evaluations, the multi-user model had worse performance than a single-user model on both speech recognition and speaker verification tasks. This is undesirable as we do not wish to degrade performance on devices with a single enrolled user. Second, the original multi-user model seems to overfit the training data and failed to generalize well to unseen combinations of enrolled users. These limitations raise severe concerns regarding the deployment of the multi-user VoiceFilter-Lite model in production environments.

In this paper, we focus on addressing these limitations by exploring variations of each component of the multi-user VoiceFilter-Lite model, and developed a new version of the model that closes the performance gap on single-user evaluations. It is important to note that in this paper, we focus only on improving speaker identification in the multi-talker setting. In summary, the original contributions of this paper include:
\begin{enumerate}
    \item We introduce a dual learning rate scheduler where the \emph{AttentionNet} is independently trained with a learning rate that is an order of magnitude smaller than the \emph{VoiceFilterNet}. Experiments in Section~\ref{sec:exp_duallr} show that the dual learning rate scheduler prevents the \emph{AttentionNet} from overfitting and significantly improves model quality.
    \item We introduce feature-wise linear modulation (FiLM) ~\cite{perez2017film,o2021conformer,narayanan2021cross} as an efficient way to condition the \emph{VoiceFilterNet} on the attended embedding. Doing so reduces model size from 3.47 MB to 3.23 MB, and significantly improves performance of the model on a speaker verification task, as shown in Section~\ref{sec:exp_film}.
    \item As a complement to the original multi-user VoiceFilter-Lite paper~\cite{rikhye2021multiuser}, we carefully compared different implementations of aggregating multiple enrolled speaker embeddings into a single embedding, and confirmed that the attention mechanism is critical to the performance, as shown in Section~\ref{sec:exp_att}.
    \item Using a combination of the best practices from above, the new multi-user VoiceFilter-Lite model performs identically to the single-user model when there is only one enrolled user, and at the same time significantly reduces speaker verification EER when there are multiple enrolled users. The resulting model meets the quality bar for deployment to production environments.
\end{enumerate}

\section{Methods}

\subsection{Review of VoiceFilter-Lite}
\label{sec:review_vfl}
VoiceFilter-Lite is a targeted voice separation model for streaming, on-device automatic speech recognition (ASR)~\cite{Wang2020}, as well as text-independent speaker verification (TI-SV)~\cite{rikhye2021personalized}. It assumes that the target speaker has completed an offline enrollment process~\cite{enrollmentblog,wang2020version}, which uses a speaker recognition model to produce an aggregated embedding vector $\mat{e}$ that presents the voice characteristics of this speaker. In this work, we use the d-vector embedding~\cite{wan2018generalized} trained with the generalized end-to-end extended-set softmax loss~\cite{pelecanos2021dr} as the speaker embedding.

Let $\mat{x}^{(t)}$ be the input feature frame at time $t$ from the speech to be processed (\emph{e.g.} stacked log Mel-filterbank energies). This feature is first frame-wise concatenated with the d-vector $\mat{e}$, then fed into an LSTM network~\cite{hochreiter1997long} followed by a fully connected neural network to produce a mask $\mat{y}^{(t)}$, which has the same dimensionality as $\mat{x}^{(t)}$:
\begin{equation}
    \label{eq:voicefilterlite}
    \mat{y}^{(t)} = \mathrm{FC} \circ \mathrm{LSTM} ( \mathrm{Concat} (\mat{x}^{(t)}, \mat{e}) ) .
\end{equation}

At runtime, the mask $\mat{y}^{(t)}$ is element-wise multiplied to the input $\mat{x}^{(t)}$ to produce the final enhanced features. Separately, we also use another LSTM-based neural network followed by a fully connected layer to estimate the noise type (either overlapping or non-overlapping speech) from the input $\mat{x}^{(t)}$. This noise type prediction is then used during inference to deactivate the VoiceFilter-Lite model when the input frame contains no overlapping speech. For more details, we refer the reader to~\cite{Wang2020}.

We have previously demonstrated that the VoiceFilter-Lite model is an important component of text-independent speaker verification ~\cite{rikhye2021personalized}. In particular, in the presence of overlapping background speech (the multitalker scenario), speaker verification tends to fail. Adding VoiceFilter-Lite to the feature frontend of speaker verification helps to suppress overlapping speech, which in turn improves the accuracy of target speaker verification. This in turn helps to reduce the false rejection rate of personalized keyphrases. For more details, we refer the reader to ~\cite{rikhye2021personalized}.

\subsection{Review of multi-user VoiceFilter-Lite}
\label{sec:review_muvf}
\begin{figure}[t!]
    \includegraphics[width=\columnwidth]{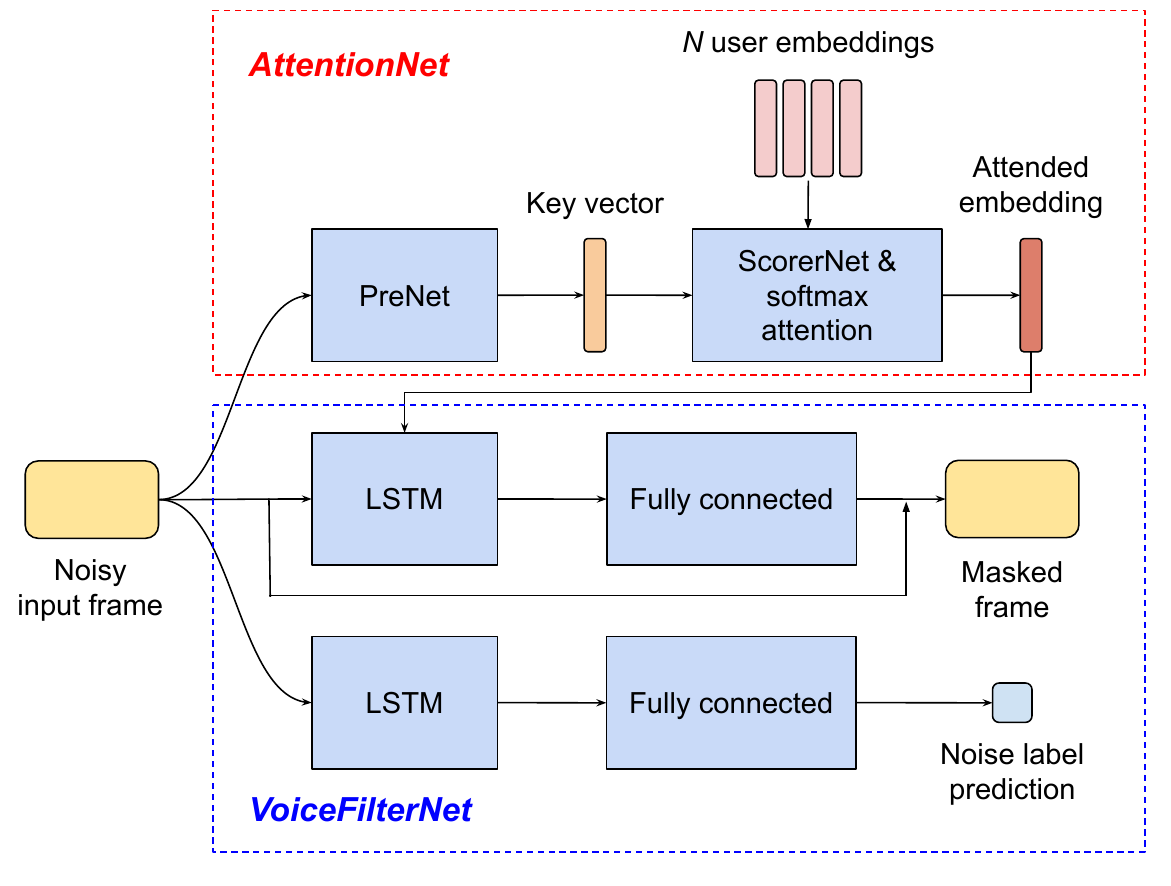}
    \caption{{\it Overall architecture of the multi-user VoiceFilter-Lite model proposed in ~\cite{rikhye2021multiuser}. This model comprises two parts --- an AttentionNet which computes the most relevant speaker from a noisy frame, and a VoiceFilterNet, which is identical to the single-use VoiceFilter-Lite model~\cite{Wang2020}.}}
    \label{fig:model-arch}
\end{figure}
%

To extend the VoiceFilter-Lite model to support multiple enrolled users, we added an \emph{AttentionNet} to the VoiceFilter-Lite model, as illustrated in Fig.~\ref{fig:model-arch}. This \emph{AttentionNet} uses an attention mechanism to compute the most relevant speaker embedding given an input frame over an inventory of multiple speaker embeddings: $\mat{e}_1, \mat{e}_2, \cdots, \mat{e}_N$. 

The \emph{AttentionNet} comprises two parts --- the \emph{PreNet} and the \emph{ScorerNet}. The \emph{PreNet} is a stack of three LSTM layers that computes, for each frame, a compressed representation of features in the stacked log Mel-filterbank energies, referred to as the \emph{key vector} $\mat{k}^{(t)}$:

\begin{equation}
    \label{eq:prenet}
    \mat{k}^{(t)} = \mathrm{PreNet}(\mat{x}^{(t)}) .
\end{equation}

This compressed representation is then individually combined with each of the $N$ enrolled speaker embeddings in the \emph{ScorerNet} to generate a score for each enrolled speaker. The attention weights $\alpha_i^{(t)} > 0$ are the softmax over these scores:

\begin{equation}
    \label{eq:scorer_net}
    s_i^{(t)} = \mathrm{ScorerNet}(\mathrm{Concat}(\mat{k}^{(t)}, \mat{e}_i)) ,
\end{equation}
\begin{equation}
    \label{eq:softmax}
    \alpha_i^{(t)} = \frac{\exp{(s_i^{(t)})}}{\sum _{j=1}^N \exp{(s_j^{(t)})}} .
\end{equation}

Finally, the attended embedding is the dot product of these attention weights and the matrix of the $N$ enrolled speaker emebddings. In this way, the \emph{ScorerNet} selects one of the $N$ enrolled speaker embeddings that is most relevant to the compressed representation, and therefore the most probable speaker in that frame:

\begin{equation}
    \label{eq:weighted_average}
    \mat{e}_\mathrm{att}^{(t)} = \sum_{i=1}^N \alpha_i^{(t)} \cdot \mat{e}_i .
\end{equation}

This attended embedding is used as a conditioning input in the \emph{VoiceFilterNet}, which is identical to the original VoiceFilter-Lite as described previously:
\begin{equation}
    \label{eq:muvf}
    \mat{y}^{(t)} = \mathrm{FC} \circ \mathrm{LSTM} ( \mathrm{Concat} (\mat{x}^{(t)}, \mat{e}_\mathrm{att}^{(t)}) ) .
\end{equation}

Both the \emph{AttentionNet} and \emph{VoiceFilterNet} in the multi-user VoiceFilter-Lite model are jointly trained  with an Adam optimizer~\cite{kingma2014adam} using a weighted linear combination of the following three loss functions:
\begin{enumerate}
    \item $L_\mathrm{asym}$: an asymmetric L2 loss  for signal reconstruction;
    \item $L_\mathrm{noise}$: a noise type prediction loss  for adaptive suppression at runtime;
    \item $L_\mathrm{att}$: an attention loss that measures how well the attention weights predict the target speaker:
    \begin{equation}
    \label{eq:att_loss}
    L_\mathrm{att}= \sum _ t \mathrm{CrossEntropy}( \mat{\alpha}^{(t)},  \mat{w}_\mathrm{gt} ) + \lambda||\mat{\alpha}^{(t)}||_{\infty} 
\end{equation}
where $\mat{w}_\mathrm{gt}$ is the ground truth attention weights and $\lambda$ is the weight of the $\mathrm{L}_\infty$ regularization term. The ground truth attention weights are the one-hot encoding of the position of the target speaker embedding. For example, if $N=4$ and the target speaker is the second enrolled speaker, then $\mat{w}_\mathrm{gt} = [{0, 1, 0, 0}]$. This loss ensures that the attention weights of the non-target speakers tend towards $0$, while the target speaker weight tends to $1$.
\end{enumerate}

For more details on the performance on this model on a variety of tasks, we refer the reader to~\cite{rikhye2021multiuser}. 

\subsection{Closing the performance gap between the single-user and multi-user models}

\begin{figure}[t!]
    \includegraphics[width=\columnwidth]{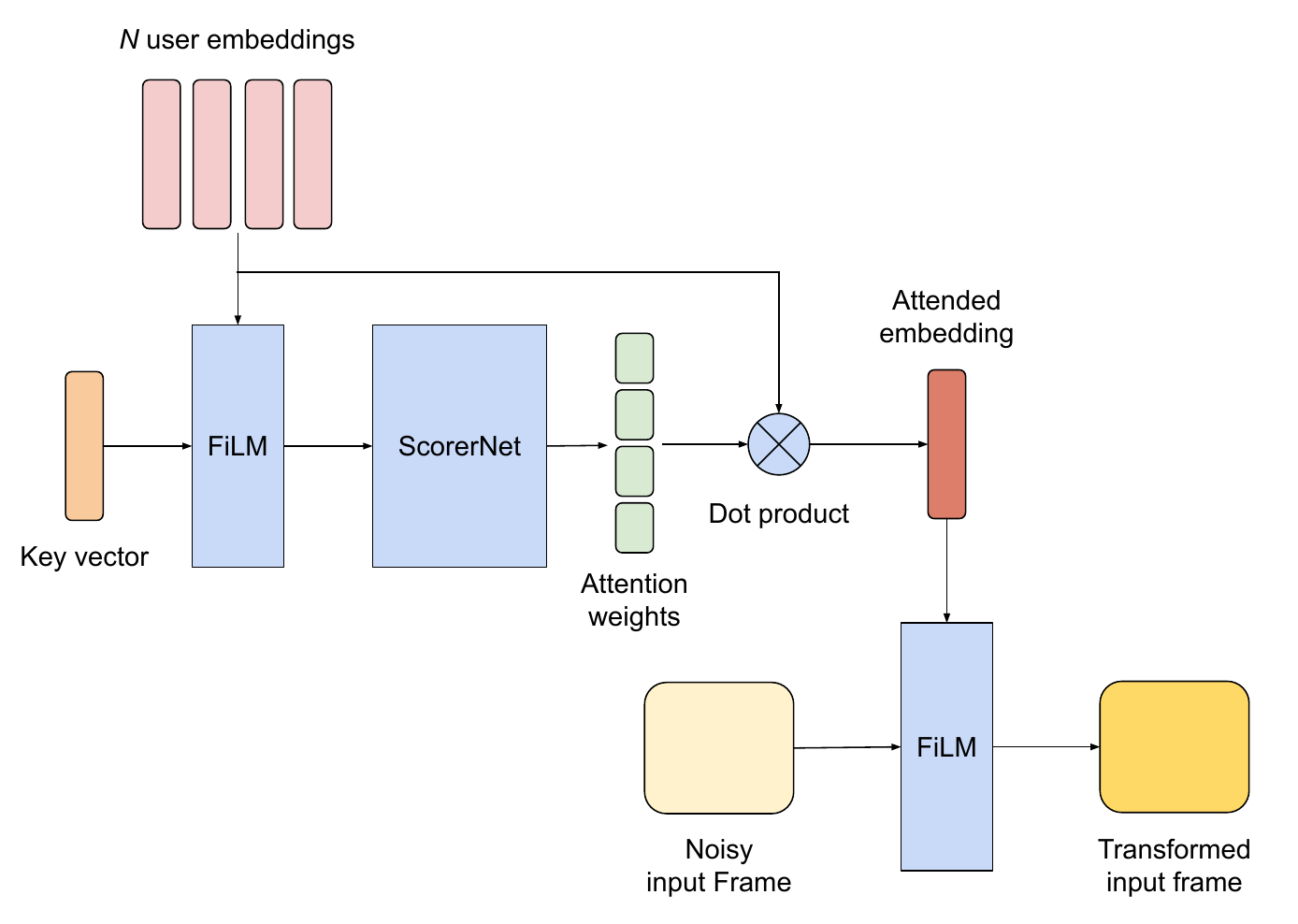}
    \caption{{\it Anatomy of the proposed new AttentionNet with FiLM-based speaker modulation.}}
    \label{fig:updated-model}
\end{figure}

As previously discussed in Section~\ref{sec:intro}, the original multi-user VoiceFilter-Lite described in Section~\ref{sec:review_muvf} suffers from performance degradation on single-user evaluations when compared with single-user VoiceFilter-Lite models, which prevents us from deploying such models in production environments. However, we observed an interesting fact --- the loss functions of the multi-user VoiceFilter-Lite model look reasonable during training. This implies the attention mechanism in the original multi-user VoiceFilter-Lite model is likely overfitting the training data, and specifically, the combinations of enrolled speakers in the training data.

To address this overfitting issue, we use a dual learning rate schedule, where the \emph{AttentionNet} is trained independently and with a smaller learning rate than the \emph{VoiceFilterNet}. Doing so ensures smaller weight updates for the \emph{AttentionNet}, allowing the optimizer to more effectively minimize the loss function to produce an optimal solution. We found that this approach prevents the \emph{AttentionNet} from memorizing the training data, which in turn allows it to generalize better to unseen examples.

To further improve the VoiceFilter-Lite model for speaker verification, we also replace the frame-wise concatenation operation between the attended embedding $\mat{e}_\mathrm{att}^{(t)}$ and the input features $\mat{x}^{(t)}$ with a feature-wise linear modulation (FiLM). In FiLM, the input features are modulated by the embedding via the following affine transformation:

\begin{equation}
    \label{eq:film}
    \mat{x}^{(t)}_\mathrm{trans} = \mathrm{FC}_{1}(\mat{e}_\mathrm{att}^{(t)}) \odot \mat{x}^{(t)} + \mathrm{FC}_{2}(\mat{e}_\mathrm{att}^{(t)}) ,
\end{equation}
\begin{equation}
    \label{eq:muvf_film}
    \mat{y}^{(t)} = \mathrm{FC} \circ \mathrm{LSTM} (\mat{x}^{(t)}_\mathrm{trans} ) ,
\end{equation}
where $\mathrm{FC}_{1}$ and $\mathrm{FC}_{2}$ are two different fully connected neural networks, and $\odot$ denotes the element-wise product. We used two-layer FC networks, where the final layer projects the attended embedding to the same dimension as the input features with a $\tanh$ activation function. Unlike concatenation, FiLM  learns to influence each input frame in an element-wise fashion by applying an affine transformation. As a result, the attended embedding is able to scale features in the input frame up or down, or negate them or even selectively threshold them allowing a more fine grained control than simple concatenation. Furthermore, FiLM only requires two parameters ($\mathrm{FC}_{1}$ and $\mathrm{FC}_{2}$) per input frame, making it a computationally more efficient conditioning method. Numerous studies have described the benefit of using FiLM in ASR~\cite{kim2017dynamic, yousefi2021speaker} and speech enhancement~\cite{narayanan2021cross,o2021conformer}, demonstrating FiLM’s broad relevance for speaker-conditioned speech models.

\section{Experimental Setup}
\label{sec:exp_setup}
\subsection{Experimental design}

Although the multi-user VoiceFilter-Lite model supports an arbitrary number of enrolled speaker embeddings as side input, there are additional constraints to consider when implementing this model in TFLite~\cite{alvarez2016efficient,shangguan2019optimizing}. Since TFLite does not support inputs with an unknown dimension, we had to pre-define a maximal number of enrolled speaker embeddings, \emph{i.e.} $N$ in our implementation. Then, at runtime, if the actual number of enrolled speakers is smaller than $N$, we use an \textbf{all-zero vector} as the embedding of any missing speaker.
Thus in the experiments to be shown in Section~\ref{sec:exp_results}, for simplicity, we first assume the maximal number of speakers is $N=2$ in our studies. Then in Section~\ref{sec:exp_4user}, we demonstrate that the observations from $N=2$ experiments are also valid when we extend it to $N=4$. 

Furthermore, in this paper, we focus only on addressing the multi-talker speaker verification challenge, especially for the \emph{multi-user multi-talker} case. For example, when both speaker $A$ and speaker $B$ enrolled their voices on the device (multi-user), and at runtime, speaker $A$ and speaker $C$ speak at the same time (multi-talker), we expect the speaker verification system to \underline{accept} the input, because it contains speech from one of the enrolled users (\emph{i.e.} speaker $A$).

For consistency, the acoustic feature frontend and the speaker verification model we used in our experiments are exactly the same as the ones used in~\cite{rikhye2021multiuser}.

\subsection{Model topology}
\label{sec:topology}
For all the models in our experiments, the \emph{VoiceFilterNet} has 3 LSTM layers, each with 256 nodes, and a fully connected layer with sigmoid activation function. The noise type prediction network has 2 LSTM layers, each with 128 nodes, and a fully connected layer with 64 nodes. In the multi-user setup, the \emph{PreNet} has 3 LSTM layers, each with 128 nodes; the \emph{ScorerNet} has two feedforward layers, each with 64 nodes.

\subsection{Training and evaluation data}
\label{sec:data}

\begin{table}
\centering
 \caption{\it Number of utterances and speakers in each subset of training and evaluation data. }
 \vspace{0.2cm}
 \label{tab:data}
\begin{tabular}{|c|r|r|}
\hline
\multicolumn{3}{|c|}{\textbf{Training data for VoiceFilter-Lite}} \\ \hline
 & Num. of utts & Num. of spks  \\ \hline
 LibriSpeech & 281,241 & 2,338 \\ 
 Vendor-collected & 2,620,867 & 16,513 \\ \hline 
 \hline
 \multicolumn{3}{|c|}{\textbf{Evaluation data for speaker verification}} \\ \hline
 & Num. of utts & Num. of spks  \\ \hline
 Enrollment set & 8,069 & 1,434  \\ 
Test set & 194,890 & 1,241  \\ 
Interference speech & 220,092 & 958  \\ \hline
\end{tabular}
\vspace{-0.3cm}
\end{table}

All the VoiceFilter-Lite models in our experiments are trained on a combination of: (1) The LibriSpeech training set~\cite{panayotov2015librispeech}; and (2) a vendor-collected dataset of English speech queries with a grand total of 2,902,102 utterances from 18,851 unique speakers (see Table~\ref{tab:data}).
To generate the noisy inputs, we augment these training utterances with different noise sources (speech and non-speech) and with different room configurations ~\cite{lippmann1987multi, ko2017study, kim2017generation}, using a signal-to-noise ratio (SNR) drawn from a uniform distribution between $1$dB and $10$dB. In the multi-user setup, each training utterance is attached with both the target speaker embedding, and randomly sampled speaker embeddings from other speakers. For example, for a 4-enrolled user model, we randomly sample 3 speaker embeddings from other speakers. And to ensure that we train on all possible speaker combinations (\emph{e.g.} 1, 2, and 3 enrolled users), we use a dropout probability of $25\%$ to randomly replace each non-target speaker embedding with an all-zero vector.

For evaluation, we use a vendor-provided English speech query dataset. The enrollment list comprises 8,069 utterances from 1,434 speakers, while the test list comprises 194,890 utterances from 1,241 speakers. Each speaker verification task is evaluated on 193k positive trials and 200k negative trials based on the enrollment and test sets. The interfering speech are drawn from a separate English dev-set consisting of 220,092 utterances from 958 speakers. More details on each subset of the training and evaluation data are provided in Table~\ref{tab:data}. During evaluation, we apply different noise sources and room configurations to the data. We use ``Clean" to denote the original non-noisified data, although they could be quite noisy already.
The non-speech noise source consists of ambient noises recorded in cafes, vehicles, and quiet environments, as well as audio clips of music and sound effects downloaded from Getty Images~\cite{getty}.
The speech noise source is a distinct development set without overlapping speakers from the testing set. We evaluate on reverberating room conditions, which consists of 3 million convolutional room impulse responses generated by a room simulator~\cite{kim2017generation} with three SNR values: $-5$dB, $0$dB, and $5$dB. 

\section{Experimental Results}
\label{sec:exp_results}

\begin{table*}[t]
\centering
 \caption{\it Equal Error Rate (EER) of text-independent speaker verification with different VoiceFilter-Lite (VF) models in the frontend and different number of enrolled users. The multi-user VoiceFilter-Lite models all use dual learning rates. Bold green text indicates best model.}
 \vspace{0.2cm}
 \label{tab:experiment1}
\begin{tabular}{|cc|c|c|ccc|ccc|}
\hline
\multicolumn{2}{|c|}{\multirow{2}{*}{\textbf{Model Name}}} & \multirow{2}{*}{\begin{tabular}[c]{@{}c@{}}\textbf{Num. of}\\ \textbf{enrolled users}\end{tabular}} & \multirow{2}{*}{\textbf{Clean}} & \multicolumn{3}{c|}{\textbf{\begin{tabular}[c]{@{}c@{}}Non-speech Noise\end{tabular}}} & \multicolumn{3}{c|}{\textbf{\begin{tabular}[c]{@{}c@{}}Speech Noise\end{tabular}}} \\ \cline{5-10} 
\multicolumn{2}{|c|}{} &  &  & \multicolumn{1}{c|}{\textbf{-5dB}} & \multicolumn{1}{c|}{\textbf{0dB}} & \textbf{5dB} & \multicolumn{1}{c|}{\textbf{-5dB}} & \multicolumn{1}{c|}{\textbf{0dB}} & \textbf{5dB} \\ \hline
\multicolumn{2}{|c|}{No VoiceFilter} & - & 0.71 & \multicolumn{1}{c|}{5.04} & \multicolumn{1}{c|}{2.23} & 1.50 & \multicolumn{1}{c|}{12.40} & \multicolumn{1}{c|}{8.29} & 5.13 \\ \hline
\multicolumn{2}{|c|}{Single-user VoiceFilter} & 1 & 0.71 & \multicolumn{1}{c|}{5.01} & \multicolumn{1}{c|}{2.19} & 1.48 & \multicolumn{1}{c|}{3.97} & \multicolumn{1}{c|}{2.42} & 1.65 \\ \hline
\multicolumn{1}{|c|}{\multirow{8}{*}{Multi-user VoiceFilter}} & \multirow{2}{*}{Averaging Model} & 1 & 0.71 & \multicolumn{1}{c|}{5.01} & \multicolumn{1}{c|}{2.20} & 1.48 & \multicolumn{1}{c|}{4.75} & \multicolumn{1}{c|}{2.66} & 1.72 \\ \cline{3-10} 
\multicolumn{1}{|c|}{} &  & 2 & 0.71 & \multicolumn{1}{c|}{5.02} & \multicolumn{1}{c|}{2.21} & 1.48 & \multicolumn{1}{c|}{7.12} & \multicolumn{1}{c|}{3.91} & 2.29 \\ \cline{2-10} 
\multicolumn{1}{|c|}{} & \multirow{2}{*}{Concat Model} & 1 & 0.71 & \multicolumn{1}{c|}{5.01} & \multicolumn{1}{c|}{2.20} & 1.48 & \multicolumn{1}{c|}{4.57} & \multicolumn{1}{c|}{2.58} & 1.72 \\ \cline{3-10} 
\multicolumn{1}{|c|}{} &  & 2 & 0.71 & \multicolumn{1}{c|}{5.02} & \multicolumn{1}{c|}{2.21} & 1.48 & \multicolumn{1}{c|}{7.41} & \multicolumn{1}{c|}{3.98} & 2.29 \\ \cline{2-10} 
\multicolumn{1}{|c|}{} & \multirow{2}{*}{\begin{tabular}[c]{@{}c@{}}\textgreen{AttentionNet} \\ \textgreen{+ Weighted Sum Model}\end{tabular}} & 1 & 0.71 & \multicolumn{1}{c|}{5.01} & \multicolumn{1}{c|}{2.22} & 1.49 & \multicolumn{1}{c|}{\textgreen{3.94}} & \multicolumn{1}{c|}{\textgreen{2.37}} & \textgreen{1.64} \\ \cline{3-10} 
\multicolumn{1}{|c|}{} &  & 2 & 0.72 & \multicolumn{1}{c|}{5.03} & \multicolumn{1}{c|}{2.21} & 1.47 & \multicolumn{1}{c|}{\textgreen{7.11}} & \multicolumn{1}{c|}{\textgreen{3.55}} & \textgreen{1.99} \\ \cline{2-10} 
\multicolumn{1}{|c|}{} & \multirow{2}{*}{\begin{tabular}[c]{@{}c@{}}AttentionNet \\ + Concat Top-K Model\end{tabular}} & 1 & 0.71 & \multicolumn{1}{c|}{5.01} & \multicolumn{1}{c|}{2.20} & 1.48 & \multicolumn{1}{c|}{3.92} & \multicolumn{1}{c|}{2.39} & 1.62 \\ \cline{3-10} 
\multicolumn{1}{|c|}{} &  & 2 & 0.72 & \multicolumn{1}{c|}{5.02} & \multicolumn{1}{c|}{2.22} & 1.49 & \multicolumn{1}{c|}{7.23} & \multicolumn{1}{c|}{3.77} & 2.13 \\ \hline
\end{tabular}
\vspace{-0.3cm}
\end{table*}

\subsection{Experiment 1 - Attention is required for accurate voice separation}
\label{sec:exp_att}
The aim of our first experiment is to determine whether the \emph{AttentionNet} is required or not. There are two naive alternative approaches one can feed the $N$ enrolled speaker embeddings to the VoiceFilter-Lite model without the \emph{AttentionNet}:
\begin{itemize}
    \item \textbf{Averaging Model}: The attended embedding is the average (arithmetic mean) of all enrolled speaker embeddings.
    \item \textbf{Concat Model}: The attended embedding is an \emph{unordered} concatenation of all enrolled speakers embeddings. To preserve the size of the attended embeddig, we linearly project this concatenated vector down to the size of a single speaker embedding.
\end{itemize}

We also explore two variations of the attention-based model, where we generate the attended embedding by:
\begin{itemize}
    \item \textbf{Weighted Sum Model}: We take the dot product (see Eq.~\ref{eq:weighted_average}) between the attention weights and the $N$ speaker embeddings, to compute a weighted sum of all the enrolled speaker embeddings. This is different from the \emph{Averaging Model}, which uses identical weight $\frac{1}{N}$ for each enrolled speaker.
    \item \textbf{Concat Top-K Model}: Out of $N$ enrolled speaker embeddings, we pick the $K$ embeddings with the largest attention weights, and concatenate them by the order of the corresponding attention weights, and project this vector down to the size of a single speaker embedding. It is important to note that this is different from the naive \emph{Concat Model},  because here the concatenated speaker embeddings are ordered by their attention weights.
\end{itemize}

The evaluation results are shown in Table~\ref{tab:experiment1}. From this table, we make three key observations. First, compared to ``No VoiceFilter", adding VoiceFilter-Lite model (rel. $-70.8\%$ for single-user and $-71.4\%$ for the best multi-user model for speech noise at SNR $0$ dB) to the feature frontend of the text-independent speaker verification system significantly reduces the equal error rate for speaker verification, confirming our previous results ~\cite{rikhye2021personalized, rikhye2021multiuser}. Since VoiceFilter-Lite is disabled when there is no overlapping speech, we observe no difference in the EER for the non-speech noise cases for all models.

Second, amongst the multi-user VoiceFilter-Lite models, we see that neither the Averaging Model (relative $10.1\%$ increase in EER for two enrolled users) nor the Concat Model (rel. $12.1\%$ increase in EER) perform as well as the attention-based multi-user VoiceFilter-Lite models. This result suggests that the \emph{AttentionNet}, which finds the most relevant target speaker, is required for good performance. Importantly, the \emph{AttentionNet} is also able to find the target speaker from evaluation data, which it has not seen during training. Furthermore, since the multi-user VoiceFilter-Lite models with \emph{AttentionNet} have single user EERs that closely match the single-user VoiceFilter-Lite model, we can confidently say that the attention mechanism is indeed able to generalize to unseen examples and is able to correctly identify the target speaker. 

Third, between the two \emph{AttentionNet} models, we find that the Weighted Sum Model outperforms the Concat Top-K Model for the two-enrolled speaker case (rel. $6.2\%$ increase in EER). Similarly, we notice that the Averaging Model also performs better than the Concat Model for the same two-enrolled speaker case. This suggest that concatenating the two speaker embeddings, with or without ordering, and then projecting it to 256 dimensions (i.e. the size of a single speaker embedding) does not contain sufficient information for the \emph{VoiceFilterNet} to identify and enhance speech features of the target speaker in the input data. Rather, using a weighted sum of the speaker embeddings is a much better predictor of the target speaker embedding. The difference in single-user EER between the Averaging Model and the Weighted Sum model further reinforces the fact that the \emph{AttentionNet} is selecting the correct speaker. 

Taken together, the results of our first experiment indicate that the \emph{AttentionNet} with weighted sum is critical to the multi-user VoiceFilter-Lite model. The simpler, non-attention-based strategies are insufficient for such tasks. In all multi-user VoiceFilter-Lite models in subsequent sections, we will use the \emph{AttentionNet + Weighted Sum Model} configuration.

\begin{table}
\centering
\caption{\it EER of text-independent speaker verification with different VoiceFilter-Lite (VFL) models. Here, we vary the learning rate (LR). Each model is trained for 25 million steps. All models use the weighted sum attention mechanism. ``Num. Spk" is the number of enrolled speakers during evaluation. For the ``Dual LR" setup, we use a LR of $10^{-5}$ for VoiceFilterNet and a LR of $10^{-6}$ for AttentionNet.}
 \vspace{0.2cm}
 \label{tab:experiment2}
\begin{tabular}{|c|c|c|ccc|}
\hline
\multirow{2}{*}{\textbf{Model Name}} & \multirow{2}{*}{\textbf{\begin{tabular}[c]{@{}c@{}}Num. \\ users\end{tabular}}} & \multirow{2}{*}{\textbf{Clean}} & \multicolumn{3}{c|}{\textbf{\begin{tabular}[c]{@{}c@{}}Speech Noise\end{tabular}}} \\ \cline{4-6} 
 &  &  & \multicolumn{1}{c|}{\textbf{-5dB}} & \multicolumn{1}{c|}{\textbf{0dB}} & \textbf{5dB} \\ \hline
No VFL & - & 0.71 & \multicolumn{1}{c|}{12.40} & \multicolumn{1}{c|}{8.29} & 5.13 \\ \hline
\begin{tabular}[c]{@{}c@{}}Single-user VFL\\ LR: $10^{-5}$\end{tabular} & 1 & 0.71 & \multicolumn{1}{c|}{3.97} & \multicolumn{1}{c|}{2.42} & 1.65 \\ \hline
\begin{tabular}[c]{@{}c@{}}Single-user VFL\\ LR: $10^{-6}$\end{tabular} & 1 & 0.71 & \multicolumn{1}{c|}{6.67} & \multicolumn{1}{c|}{3.79} & 2.24 \\ \hline
\multirow{2}{*}{\begin{tabular}[c]{@{}c@{}}Multi-user VFL\\ LR: $10^{-5}$\end{tabular}} & 1 & 0.71 & \multicolumn{1}{c|}{4.13} & \multicolumn{1}{c|}{2.50} & 1.68 \\ \cline{2-6} 
 & 2 & 0.71 & \multicolumn{1}{c|}{10.39} & \multicolumn{1}{c|}{6.79} & 4.31 \\ \hline
\multirow{2}{*}{\begin{tabular}[c]{@{}c@{}}Multi-user VFL\\ LR: $10^{-6}$\end{tabular}} & 1 & 0.71 & \multicolumn{1}{c|}{6.97} & \multicolumn{1}{c|}{3.88} & 2.25 \\ \cline{2-6} 
 & 2 & 0.71 & \multicolumn{1}{c|}{9.52} & \multicolumn{1}{c|}{5.24} & 2.81 \\ \hline
\multirow{2}{*}{\textgreen{\begin{tabular}[c]{@{}c@{}}Multi-user VFL\\ Dual LR\end{tabular}}} & 1 & 0.71 & \multicolumn{1}{c|}{\textgreen{4.02}} & \multicolumn{1}{c|}{\textgreen{2.51}} & \textgreen{1.73} \\ \cline{2-6} 
 & 2 & 0.71 & \multicolumn{1}{c|}{\textgreen{8.46}} & \multicolumn{1}{c|}{\textgreen{4.84}} & \textgreen{3.43} \\ \hline
\end{tabular}
\vspace{-0.3cm}
\end{table}

\subsection{Experiment 2 - Dual learning rate schedule helps to avoid AttentionNet overfitting}
\label{sec:exp_duallr}

One observation we made in our previous multi-user VoiceFilter-Lite study~\cite{rikhye2021multiuser} is that the attention mechanism tends to overfit and memorize training data. Our next experiment is aimed at addressing this limitation by tuning the learning rate of the model.

Evaluation results are shown in Table~\ref{tab:experiment2}. Since changing the learning rate or model architecture does not affect performance on non-speech background noise (see Table~\ref{tab:experiment1}), we omit the non-speech noise results from the next two tables. 

First, for the single-user VoiceFilter-Lite model, we notice that using a smaller learning rate of $10^{-6}$ results in a significantly worse model with a much higher EER across all SNR values compared to the model trained with a higher learning rate $10^{-5}$ (rel. $56.6\%$ increase at SNR 0 dB). Secondly, for the multi-user VoiceFilter-Lite model, we observe a regression in the EER with two-enrolled users with the higher learning rate (rel. $29.6\%$ increase at SNR 0 dB). This suggests that with a higher learning rate the \emph{AttentionNet} tends to overfit on training data and fails to generalize to the evaluation data. Therefore, we implemented a dual learning rate scheduler where the \emph{AttentionNet} is trained with a smaller learning rate of $10^{-6}$, while the \emph{VoiceFilterNet} is trained with a larger learning rate of $10^{-5}$. As shown in Table~\ref{tab:experiment2}, this significantly improves both the single- and two-user ($28.7\%$ reduction relative to the original learning rate at SNR 0 dB) performance of the model.

\begin{table}
\centering
 \caption{\it EER of text-independent speaker verification with different VoiceFilter-Lite (VFL) models. Here, the attended embedding conditioning mechanism is changed. All multi-user VFL models use Weighted Sum and Dual Learning Rate schedule. Bold green text indicates best model. }
 \vspace{0.2cm}
 \label{tab:experiment3}
\begin{tabular}{|c|c|c|ccc|}
\hline
\multirow{2}{*}{\textbf{Model Name}} & \multirow{2}{*}{\textbf{\begin{tabular}[c]{@{}c@{}}Num. \\ users\end{tabular}}} & \multirow{2}{*}{\textbf{Clean}} & \multicolumn{3}{c|}{\textbf{\begin{tabular}[c]{@{}c@{}}Speech Noise\end{tabular}}} \\ \cline{4-6} 
 &  &  & \multicolumn{1}{c|}{\textbf{-5dB}} & \multicolumn{1}{c|}{\textbf{0dB}} & \textbf{5dB} \\ \hline
No VFL & - & 0.71 & \multicolumn{1}{c|}{12.40} & \multicolumn{1}{c|}{8.29} & 5.13 \\ \hline
Single-user VFL & 1 & 0.71 & \multicolumn{1}{c|}{3.97} & \multicolumn{1}{c|}{2.42} & 1.65 \\ \hline
\multirow{2}{*}{\begin{tabular}[c]{@{}c@{}}Multi-user VFL \\ + Concat Cond. \end{tabular}} & 1 & 0.71 & \multicolumn{1}{c|}{4.02} & \multicolumn{1}{c|}{2.51} & 1.73 \\ \cline{2-6} 
 & 2 & 0.71 & \multicolumn{1}{c|}{8.46} & \multicolumn{1}{c|}{4.84} & 3.49 \\ \hline
 \multirow{2}{*}{\textgreen{\begin{tabular}[c]{@{}c@{}}Multi-user VFL \\ + FiLM Cond. \end{tabular}}} & 1 & 0.71 & \multicolumn{1}{c|}{\textgreen{3.94}} & \multicolumn{1}{c|}{\textgreen{2.37}} & \textgreen{1.64} \\ \cline{2-6} 
 & 2 & 0.71 & \multicolumn{1}{c|}{\textgreen{7.11}} & \multicolumn{1}{c|}{\textgreen{3.55}} & \textgreen{1.99} \\ \hline
\end{tabular}
\vspace{-0.3cm}
\end{table}

\begin{table*}[t]
\centering
\caption{\it EER of text-independent speaker verification with different VoiceFilter-Lite (VFL) models. Here, the best 2-enrolled user and best 4-enrolled user are compared with the previously published model.}
 \vspace{0.2cm}
 \label{tab:experiment4}
\begin{tabular}{|c|c|c|ccc|ccc|}
\hline
\multirow{2}{*}{\textbf{Model Name}} & \multirow{2}{*}{\textbf{\begin{tabular}[c]{@{}c@{}}Num. of\\ enrolled users\end{tabular}}} & \multirow{2}{*}{\textbf{Clean}} & \multicolumn{3}{c|}{\textbf{\begin{tabular}[c]{@{}c@{}}Non-speech Noise\end{tabular}}} & \multicolumn{3}{c|}{\textbf{\begin{tabular}[c]{@{}c@{}}Speech Noise\end{tabular}}} \\ \cline{4-9} 
 &  &  & \multicolumn{1}{c|}{\textbf{-5dB}} & \multicolumn{1}{c|}{\textbf{0dB}} & \textbf{5dB} & \multicolumn{1}{c|}{\textbf{-5dB}} & \multicolumn{1}{c|}{\textbf{0dB}} & \textbf{5dB} \\ \hline
No VFL & - & 0.71 & \multicolumn{1}{c|}{5.04} & \multicolumn{1}{c|}{2.23} & 1.50 & \multicolumn{1}{c|}{12.40} & \multicolumn{1}{c|}{8.29} & 5.13 \\ \hline
Single-user VFL & 1 & 0.71 & \multicolumn{1}{c|}{5.01} & \multicolumn{1}{c|}{2.19} & 1.48 & \multicolumn{1}{c|}{3.97} & \multicolumn{1}{c|}{2.42} & 1.65 \\ \hline
\multirow{2}{*}{\begin{tabular}[c]{@{}c@{}}Best \\ Two-user VFL\end{tabular}} & 1 & 0.71 & \multicolumn{1}{c|}{5.01} & \multicolumn{1}{c|}{2.22} & 1.49 & \multicolumn{1}{c|}{3.94} & \multicolumn{1}{c|}{2.37} & 1.64 \\ \cline{2-9} 
 & 2 & 0.72 & \multicolumn{1}{c|}{5.03} & \multicolumn{1}{c|}{2.21} & 1.47 & \multicolumn{1}{c|}{7.11} & \multicolumn{1}{c|}{3.55} & 1.99 \\ \hline
\multirow{4}{*}{\begin{tabular}[c]{@{}c@{}}Previously Published \\ Four-user VFL ~\cite{rikhye2021multiuser}\end{tabular}} & 1 & 0.71 & \multicolumn{1}{c|}{5.03} & \multicolumn{1}{c|}{2.21} & 1.47 & \multicolumn{1}{c|}{7.32} & \multicolumn{1}{c|}{3.90} & 2.19 \\ \cline{2-9} 
 & 2 & 0.72 & \multicolumn{1}{c|}{5.04} & \multicolumn{1}{c|}{2.21} & 1.49 & \multicolumn{1}{c|}{9.34} & \multicolumn{1}{c|}{5.18} & 2.78 \\ \cline{2-9} 
 & 3 & 0.72 & \multicolumn{1}{c|}{5.01} & \multicolumn{1}{c|}{2.22} & 1.49 & \multicolumn{1}{c|}{10.36} & \multicolumn{1}{c|}{5.73} & 3.01 \\ \cline{2-9} 
 & 4 & 0.72 & \multicolumn{1}{c|}{5.05} & \multicolumn{1}{c|}{2.21} & 1.49 & \multicolumn{1}{c|}{10.99} & \multicolumn{1}{c|}{6.10} & 3.14 \\ \hline
\multirow{4}{*}{\begin{tabular}[c]{@{}c@{}}New\\ Four-user VFL\end{tabular}} & 1 & 0.71 & \multicolumn{1}{c|}{5.03} & \multicolumn{1}{c|}{2.21} & 1.47 & \multicolumn{1}{c|}{4.32} & \multicolumn{1}{c|}{2.54} & 1.71 \\ \cline{2-9} 
 & 2 & 0.71 & \multicolumn{1}{c|}{5.03} & \multicolumn{1}{c|}{2.21} & 1.47 & \multicolumn{1}{c|}{7.59} & \multicolumn{1}{c|}{4.21} & 2.50 \\ \cline{2-9} 
 & 3 & 0.72 & \multicolumn{1}{c|}{5.03} & \multicolumn{1}{c|}{2.22} & 1.49 & \multicolumn{1}{c|}{8.05} & \multicolumn{1}{c|}{5.14} & 2.85 \\ \cline{2-9} 
 & 4 & 0.72 & \multicolumn{1}{c|}{5.03} & \multicolumn{1}{c|}{2.21} & 1.50 & \multicolumn{1}{c|}{9.78} & \multicolumn{1}{c|}{5.38} & 2.99 \\ \hline
\end{tabular}
\vspace{-0.3cm}
\end{table*}

\subsection{Experiment 3 - FiLM-based speaker conditioning improves model performance}
\label{sec:exp_film}

So far, we have shown that having an \emph{AttentionNet} and training it with a smaller learning rate than the \emph{VoiceFilterNet} is necessary for good performance in reducing EER when the multi-user VoiceFilter-Lite model is present in the text-independent speaker verification frontend. Another aspect of the model that can be further optimized is how the attended embedding is used by the \emph{VoiceFilterNet}. 

There are several ways in which the attended embedding can be used to condition the \emph{VoiceFilterNet}:
\begin{itemize}
    \item \textbf{Concat-Conditioned Model}: The attended embedding is concatenated with each input frame before being fed into the \emph{VoiceFilterNet} LSTM stack. This increases the dimensions of the input frame by the size of the attended embedding (256 dimensions).
    \item \textbf{FiLM-Conditioned Model}: An affine transformation, shown in Eq. ~\ref{eq:film}, is applied to each input frame. This affine transformation allows the attended embedding to modulate the input frame in a feature-wise manner. This does not change the dimensions of the input frame.
\end{itemize}

Evaluation results for these different models are shown in Table~\ref{tab:experiment3}. In these experiments, we keep the \emph{AttentionNet} architecture the same (Weighted Sum Model) and use a dual learning rate schedule as we have described in the preceeding sections. We observed that the multi-user VoiceFilter-Lite model that uses FiLM to condition the input frames with the attended emebedding performs significantly better than the model that uses concatenation. Specifically, relative to the concat-conditioned multi-user model, we find a $5.6\%$ reduction in EER for one enrolled user and a $26.7\%$ reduction (at SNR 0dB) for two enrolled users. We also observe that relative to the single-user VFL, the FiLM-conditioned multi-user model has comparable performance with a single enrolled user ($2.42$ vs. $2.37$ at SNR 0dB). Therefore, using FiLM to condition the model on the speaker embedding is far more robust method than concatenation.

\subsection{Experiment 4 - Same observations hold for four enrolled users}
\label{sec:exp_4user}

Finally, we demonstrate that the best two-user model can be easily extended to support four enrolled users. In the following experiments, we trained a four-user model with the same model architecture. Evaluation results for this model is shown in Table~\ref{tab:experiment4}.

Compared to our previously published model~\cite{rikhye2021multiuser}, the new four-user model, which uses FiLM and dual learning rates (see Fig.~\ref{fig:updated-model}) results in a significantly lower EER for all speaker combinations. Interestingly, we observe a regression in EER between the best two-user VoiceFilter-Lite model and the four-user model for the 1-speaker ($2.37$ vs. $2.54$, $7.1\%$ increase at SNR 0dB) and 2-speaker ($3.55$ vs. $4.22$, $18.9\%$ increase at SNR 0dB) evaluations. One reason for this could be that there are fewer 1-speaker and 2-speaker examples during training the four-user model ($\mathrm{P}_\mathrm{1-user} = 0.25^3$) than the two-user model ($\mathrm{P}_\mathrm{1-user} = 0.25$) due to the way we process our training data (see Section~\ref{sec:data}). In fact, for the four-user model, only about $15.6\%$ of the training data contains one or two enrolled users. As a result, during evaluation, the model does not generalize very well on 1-speaker or 2-speaker evaluations compared with 3-speaker and 4-speaker evaluations. To address this issue, one of our future work directions is to balance our training data according to the realistic distributions of the number of users on shared devices, as well as to make the model more robust to unbalanced data.

\section{Conclusions}
In this paper, we devised a series of experiments to evaluate the impact of various design choices in the multi-user VoiceFilter-Lite model. We confirmed that an attention mechanism is critical for the multi-user model to function well, which cannot be replaced by naive aggregation logic such as either averaging or concatenating all enrolled speaker embeddings.
We found that training the attention mechanism with a learning rate that is an order of magnitude smaller than the rest of the model addresses the overfitting issue, and is critical to close the performance gap between single-user and multi-user models on single-user evaluations.
Additionally, the performance of the model could be further improved by using FiLM to modulate the attended speaker embedding.

Although all experiments in this paper are carried out for multi-user VoiceFilter-Lite, it is important to note that the proposed attention-based speaker selection mechanism is a generic solution that can be applied to any speaker-conditioned speech models. This is crucial as most smart home devices, such as smart displays and smart speakers, usually support multiple enrolled users.
Thus as our future work, we would like to adopt the best practices from the multi-user VoiceFilter-Lite to other speaker-conditioned speech models, including personalized ASR or personal VAD.

\bibliographystyle{IEEEbib}
\bibliography{Odyssey2022_BibEntries}

%

\end{document}